\documentstyle[prl,aps]{revtex}
\tightenlines

\begin{document}

\draft
\title{
Langevin Simulation of Nonlocal Ginzburg-Landau Model
for Superconductors in a Magnetic Field
}

\author{Ayumi FUJITA} 
\address{RIKEN (The Institute of Physical and Chemical Research)\\
Wako, SAITAMA, 351-01, JAPAN}
\date{\today}
\maketitle
\begin{abstract}
We numerically investigate the phenomenological nonlocal Ginzburg-Landau 
Hamiltonian for two-dimensional superconductors in a strong magnetic field 
by Langevin equation. We obtain a regular vortex lattice which is very near 
to the square lattice. We calculate the  structure factor and Abrikosov
factor at various temperatures.  We also evaluate the specific heat 
and obtain a cusp which indicates the melting of the vortex lattice.
\end{abstract}


The study of the high $T_c$ superconductor
(HTSC) in a magnetic field has renewed the interests in the mixed state.
The most characteristic behavior of this unconventional superconductor
is the large superconducting fluctuation, and it is known that 
the vortex liquid phase appears  below the $H_{c2}$ line in the $H-T$ 
phase diagram.
The melting transition of the vortex lattice has been studied theoretically
ever since the discovery of the HTSC.~\cite{rev,nelson}  
For the two-dimensional case, there has been a large amount of studies 
which investigate the vortex lattice melting
transition of the Kosterlitz-Thouless type which is accompanied by the 
dissociation of the dislocation pairs.~\cite{doni,fisher}
Experimentally, the vortex lattice melting transition is observed as the 
first order phase 
transition by revealing a sharp step of the magnetization in both 
$Bi_2 Sr_2 Ca Cu_2 O_{8+\delta}$ and $YBa_2 Cu_3 O_{7-\delta}$.
~\cite{zeldov,welp}  
The first order melting transition of the vortex lattice is already predicted 
theoretically from the renormalization group analysis based on 
Ginzburg-Landau model for three dimensional system.~\cite{brezin} 
Although it is not yet certainly clear whether
the first order melting transition survive in the strong magnetic
field  phase where the fluctuation effect is very large and the system behaves 
more like two-dimensional. 

On the other hand, the oxide high $T_c$ superconductor has a 
layered structure and it is considered as a 
quasi-two-dimensional system.  It is pointed out that  
the two-dimensional square lattice structure of $CuO_2$ plane has an
effect not only on the symmetry of the order parameter but also on the vortex
lattice configuration.\cite{maki}   Recently many experimental evidences 
which support the notion that the order parameter of HTSC has
$d_{x^2 -y^2}$ symmetry are reported.  The expansion of the GL model
into a form which deals with the $d$-wave superconductors are thus
necessary in order to investigate the vortex state in high-$T_c$ cuprates.

In Ref.~\cite{HF}, we have previously investigated a nonlocal GL model 
starting with following two motivations.  
One is that the nonlocal GL model has a remarkable 
similarity to the $n$-component GL model.  We find that a 
parameter of the characteristic range of the nonlocal interaction
plays a role of $n$ for the $n$-component GL model.  
The other one is that because 
of the shortness of the coherence length of HTSC, modification of the
quartic interaction of the order parameter into a nonlocal form
is considered to be reasonable.

We introduce a phenomenological nonlocal Hamiltonian,
\begin{equation}
H[\psi] = \int d^d {\bf r}\left[ \alpha|\psi|^2 + {\hbar\over 2m}|(-i\nabla
-2e{\bf A})\psi |^2 \right] + {\beta\over 2}
\int d^d {\bf r}_1 d^d {\bf r}_2 |\psi(
{\bf r}_1 ) |^2 g({\bf r}_1, {\bf r}_2 ) |\psi({\bf r}_2 )|^2 ,
\label{H}
\end{equation}
where the kernel $g({\bf r}_1 ,{\bf r}_2 )$ represents a nonlocal interaction
between $|\psi|^2$.
For the symmetric case, we choose
\begin{equation}
g({\bf r}_1 ,{\bf r}_2 ) ={1\over \pi d^2}\exp\left( -{{|{\bf r}_1 -
{\bf r}_2 |^2}\over d^2 }\right)
\end{equation}
where $d$ indicates the range of the nonlocal interaction.  In Ref.~\cite{HF}
the high temperature series for the free energy is obtained with this Gaussian 
form nonlocal interaction.

In this study, we investigate numerically the model Hamiltonian Eq.~(\ref{H}) 
in the case that the kernel
$g({\bf r}_1 ,{\bf r}_2 )$ has a four-fold symmetry with Langevin equation.  
This model is studied recently in Ref.~\cite{yeo} 
and they calculate the structure 
factor using the approximation which is valid at moderately low temperatures.  
We take the same notation as Ref.~\cite{yeo}, and the Fourier transform of 
$g({\bf r})$ is written as
\begin{equation}
\tilde g({\bf k}) = \exp \left[-C \left({k\over \mu}\right)^4 (1-\epsilon \cos
4\theta)\right]
\end{equation}
where ${\bf k}=k(\cos\theta, \sin\theta)$ and $\mu^{-1} = \sqrt{1/ 2eB}$ is
the magnetic length.  The two parameters $C$ and $\epsilon$ determine 
the amount of the effect of nonlocal interaction on the system. 
We consider the two-dimensional system in a magnetic field perpendicular to
the $xy$-plane.  
We also take the unit $\hbar = c = k_B = 1 $ and the gauge ${\bf A}=(0, xB)$.

In a magnetic field, the order parameter 
$\psi({\bf r})$ is expanded with the Landau levels.  
In order to simplify the numerical
calculation, we take the lowest Landau level approximation which is valid
in a strong magnetic field. 
Then $\psi({\bf r})$ is written as
\begin{equation}
\label{psi}
\psi({\bf r})=\sum_q a_q L_y^{-1/2}( \pi /\mu^2 )^{-1/4}
 \exp\bigl[ iqy -{\mu^2 \over 2}(x-q/\mu^2 )^2 \bigr],
\end{equation}
where $q$ is the momentum in $xy$-plane, $a_q$ is a complex number and $L_y$
is the system size.  The Hamiltonian is rewritten with this $\psi$ as
\begin{eqnarray}
H&=&\sum_q \alpha_H |a_q |^2 + {\beta\over 2L_y^2}\sum_q \int d^2 {\bf k}
\exp \biggl[ -{1\over 4\mu^2}\{ (q_1 - q_3 )^2 + (q_2 - q_4 )^2 \} -
{k_x^2 \over 2\mu^2}\nonumber\\
& &-{i\over 2\mu^2}k_x (q_1 -q_2 +q_3 -q_4 ) 
-{C\over \mu^4}\{ (1-\epsilon)(k_x^4 + k_y^4 ) +2(1+3\epsilon)k_x^2 k_y^2
\} \biggr]\nonumber\\
& & \times a_{q_1}^\ast a_{q_2}^\ast a_{q_3} a_{q_4} \delta_{q_1 + q_2 , 
q_3 + q_4 }\delta_{q_1 - q_3 , k_y}
\label{Hami2}
\end{eqnarray}

In our numerical simulation study, the momentum is discretized as 
\begin{equation}
q={2\pi\over L_y}n,\quad n=1,\cdots N,
\end{equation}
where $N$ is the number of vortices in the system with length $L_x$ and 
$L_y$, 
\begin{equation}
N={\mu^2 L_x L_y \over 2\pi }.
\end{equation}

The Langevin equation for the complex order parameter $a_q =A_q + iB_q $ is
given by 
\begin{eqnarray}
   {dA_{q}\over{dt}} &=& -{1\over{2}}{\partial H
   \over{\partial A_{q}}} + \eta \\
   {dB_{q}\over {dt}} &=& -{1\over{2}}{\partial H\over{\partial B_{q}}}
     + \eta
\end{eqnarray}
where $\eta$ is the Gaussian white noise.
With these Langevin equations we can calculate a set of $a_q$ at 
each time step, and by averaging over the time steps we obtain 
the statistical average for the values of $a_q$.  
We denote the discretized time step by $\delta$.
We rewrite $A_q = A(I)$ ,$B_q =B(I)$ where $I$ indicates the discretized 
momenta $I=1,\dots,N$.  After a
time evolution of step $\delta$, $A(I),B(I)$ becomes $A'(I)$ and $B'(I)$ 
which is given by
\begin{eqnarray}
\label{eq:langevin}
   A^{\prime}(I) &=& A(I) -{\delta\over
2}(2\alpha_{H}A(I)+S1)+\sqrt{\delta}\eta
   \\
   B^{\prime}(I) &=& B(I) -{\delta\over
2}(2\alpha_{H}B(I)+S2)+\sqrt{\delta}\eta
\end{eqnarray}
where
\begin{eqnarray}
S1&=&{\beta\over 2L_y^2}
\sum_{J_1 ,J_2 ,K_1 }\biggl( A(J_1 )A(J_2 )A(I+J_1 -J_2 )-A(J_1 )B(J_2 )
B(I+J_1 -J_2 ) \nonumber\\
& & +2B(J_1 )A(J_2 )B(I+J_1 -J_2)\biggr)
\cos(2\pi K_1 (J_2 - J_1 )/N)h(J_1 ,J_2, K_1)\\
S2&=&{\beta\over 2L_y^2}
\sum_{J_1 ,J_2 ,K_1 }\biggl( B(J_1 )B(J_2 )B(I+J_1 -J_2 )-B(J_1 )A(J_2 )
A(I+J_1 -J_2 ) \nonumber\\
& & +2A(J_1 )B(J_2 )A(I+J_1 -J_2)\biggr)
\cos(2\pi K_1 (J_2 - J_1 )/N)h(J_1 ,J_2, K_1)\\
\end{eqnarray}
and
\begin{eqnarray}
h(J_1 ,J_2 , K_1 ) &=& \exp\biggl[-{\pi\over N}\left\{(I -J_2 )^2  
+K_1^2\right\} - C(1-\epsilon)\left({2\pi\over N}\right)^2 \left(K_1^4
+ K_2^4 \right)\nonumber\\
& & -2C(1+3\epsilon)\left({2\pi\over N}\right)^2 K_1^2 K_2^2 \biggr]\\
K_2 &=& I - J_2
\end{eqnarray}

We set typical time step $\delta = 0.05$ and discard first $40000$ steps
for the equilibration. We put $N=64$ vortices in the system, and the initial
condition is taken as $A(I)=B(I)=0$ for all $I$.   
We also take the periodic boundary condition in the momentum space.
\begin{equation}
\left.
\begin{array}{lll}
A(I) &=& A(mod(I,N)) \\
B(I) &=& B(mod(I,N)) 
\end{array}\right\} \qquad\mbox{for $|I|>N$}
\end{equation}
We carry out the numerical simulation for various reduced temperatures 
$y=(T-T_c )/\Delta T$ which is related to $\alpha_H$ as
\begin{equation}
\label{eq:wai}
y=\alpha_H \sqrt{2\pi/\beta\mu^2}.
\end{equation}

Fig.~1 shows a configuration of vortices at temperatures $y=-10$ and $y=-4$. 
The parameters are taken as $C=0.01$ and $\epsilon=0.5$ in this calculation.
We plot $<|\psi(x,y)|^2 >$ whose averaging is done over $1.6\times 10^5 $ 
time steps. The white part indicates the large value of $<|\psi|^2 >$ and 
the black spots correspond to the vortices. At $y=-10$ we see a regular 
rhombic vortex lattice structure instead of the triangular lattice.  At a
temperature $y=-4$, although the magnitude of $<|\psi|^2 >$ is much
lowered the rhombic configuration is still maintained.

The vortex state is probed by the density-density correlation function which
is called structure factor.
This function is defined in the momentum space as
\begin{equation}
\Delta({\bf k}) = \sum_{q_i} a_{q_1}^\ast a_{q_2}^\ast a_{q_3} a_{q_4}\exp\left[
i\mu^{-2}(q_3 -q_1 )k_x \right]
\delta_{q_4 ,q_1 + k_y}\delta_{q_1 + q_2 , q_3 +q_4}
\end{equation}
which is directly 
related to a neutron diffraction pattern.
In the vortex lattice phase
\begin{equation}
\Delta({\bf k}) = \Delta_0 \delta_{{\bf k},{\bf G}}
\end{equation}
where ${\bf G}$ is the reciprocal lattice vectors.
In the case of triangular lattice, the diffraction pattern  shows six-fold 
sharp peaks in the first Brillouin zone.  In Fig.~2 we plot 
$\tilde\Delta({\bf k}) = \Delta({\bf k})/\Delta_0$ for three different
temperatures.
In the present nonlocal GL model, the six fold symmetry of the diffraction
pattern disappears and 
a rectangular peaks  which correspond to the nearby reciprocal
lattice points for the rhombic lattice are obtained.  
The reciprocal lattice points including 
the next nearest ones form a stretched hexagon.  This configuration
almost does not change over the temperature region $-12 \ge y\ge -4$.
The height of the peaks, however,  become very low  at a temperature 
$y\sim -4$ and this behavior resembles to that of the diffraction pattern
obtained by the local GL model at a temperature just below 
the melting point.  
In high temperature region $y>-4$, the peaks are totally disappeared.
(Note that this temperature is still lower than the $T_c$ for the mean field
theory.)
The ring like diffraction pattern for the vortex liquid phase  
is not obtained in the present model.

We also calculate the Abrikosov factor
\begin{equation}
\beta_A = {<|\psi({\bf r })|^4 >\over <|\psi({\bf r})|^2 >^2 }
\end{equation}
where $<\cdots>$ denotes the spatial average.
In the vortex lattice state $\beta_A $ is given by
\begin{equation}
\beta_A = \sum \tilde\Delta ({\bf G}) \exp (-{\bf G}^2 /2\mu^2 )
\end{equation}
where the summation is taken over the reciprocal lattice vectors ${\bf G}$.
For the triangular lattice $\beta_A = 1.16$ and for the square lattice 
$\beta_A = 1.18$.  At temperature $y=-12$ we obtain
$\beta_A = 1.20$ which is larger than the value for the square lattice.
For the rhombic lattice $\beta_A $ is evaluated as
\begin{equation}
\beta_A = \sum_{n,m} \exp\left[ -{\pi\over \cos a}\{ m^2 \cos^2 a 
+(n-m\sin^2 a)^2\}\right]
\end{equation}
where $a$ is the half of the angle which is made by two adjacent sides.
For the case of $\beta_A = 1.20$, this angle is evaluated as $a=45.84^\circ$
which is very near to the value for the square lattice.
Above the temperature $y \stackrel{>}{\sim} -6$, Abrikosov factor
grows rapidly at about $y=-4$ and it converges to the value $\beta_A = 2.0$ 
which corresponds to the uncorrelated vortex liquid phase or the normal phase.
From this behavior we see that in the nonlocal GL
model the melting transition of the vortex lattice takes place at a 
temperature near $y=-4$.

Next we calculate the specific heat.  The specific heat is obtained by taking
a second derivative of Eq.~(\ref{Hami2}) with respect to the temperature
$\alpha_H$.  
\begin{equation}
C = \left< \left({\partial H\over \partial \alpha_H}\right)^2 \right> -
\left< {\partial H\over \partial \alpha_H} \right>^2
\end{equation}
where
\begin{equation}
\left< \left({\partial H\over \partial \alpha_H}\right)^2 \right>
 = \left< \Delta(0)\right> \ ,\qquad
\left< {\partial H\over \partial \alpha_H} \right>^2 =
{1\over N}\left< \sum_q |a_q |^2 \right > 
\end{equation} 
In Fig.3 we plot the scaling function for the  specific heat 
$C/\Delta C$ where $\Delta C$ is the specific heat jump at $T_c$ in the
mean field theory.  We obtain a cusp in the scaling function of the specific 
heat at $y=-4$ at which temperature the Abrikosov factor shows 
a discontinuous jump. This singularity of the specific
heat curve has been previously obtained by the matrix GL model approach with 
large-$N$ limit where the free energy is calculated up to 
$O(1/N)$.~\cite{matrixfl}  In the matrix GL model for the large $N$ limit, 
the specific heat
curve shows remarkable agreement with the usual GL model both in low 
temperature and high temperature region.  
In the current model, however, such agreement is lost except the high
temperature region.

In summary, we studied a phenomenological nonlocal GL model numerically
using the Langevin equations.  We obtained the rhombic vortex lattice in the
low temperature phase and the corresponding value for the Abrikosov factor
$\beta_A = 1.20$.
The cusp in the specific heat suggests that in the nonlocal GL model 
the superconducting transition reveals a totally different feature  
from that in the original GL model which shows no singurality at the 
critical point from the high temperature expansion series study.\cite{HFL}
In order to determine whether the melting transition is discontinuous, 
detailed studies both with numerical and analytical methods are
necessary.

The numerical calculations were made by using the Vector Parallel Processor,
Fujitsu VPP500 at RIKEN.

\begin{figure}
\caption{The spatial distribution of  $<|\psi(x,y)|^2 >$ for the system of 
$N=64$ vortices. The black spots correspond to the vortices.}
\end{figure}

\begin{figure}
\caption{The diffraction pattern at three diffent temperatrues. We plot
only the nearby peaks from the origin ${\bf k}=0$.}
\end{figure}

\begin{figure}
\caption{The scaling function of the specific heat $C/\Delta C$.  
The results from the different iteration times are shown for the check of 
equilibration at various temperatures. 
The cusp at $y=-4$ is enhanced by taking a longer  
iteration time.}
\end{figure}

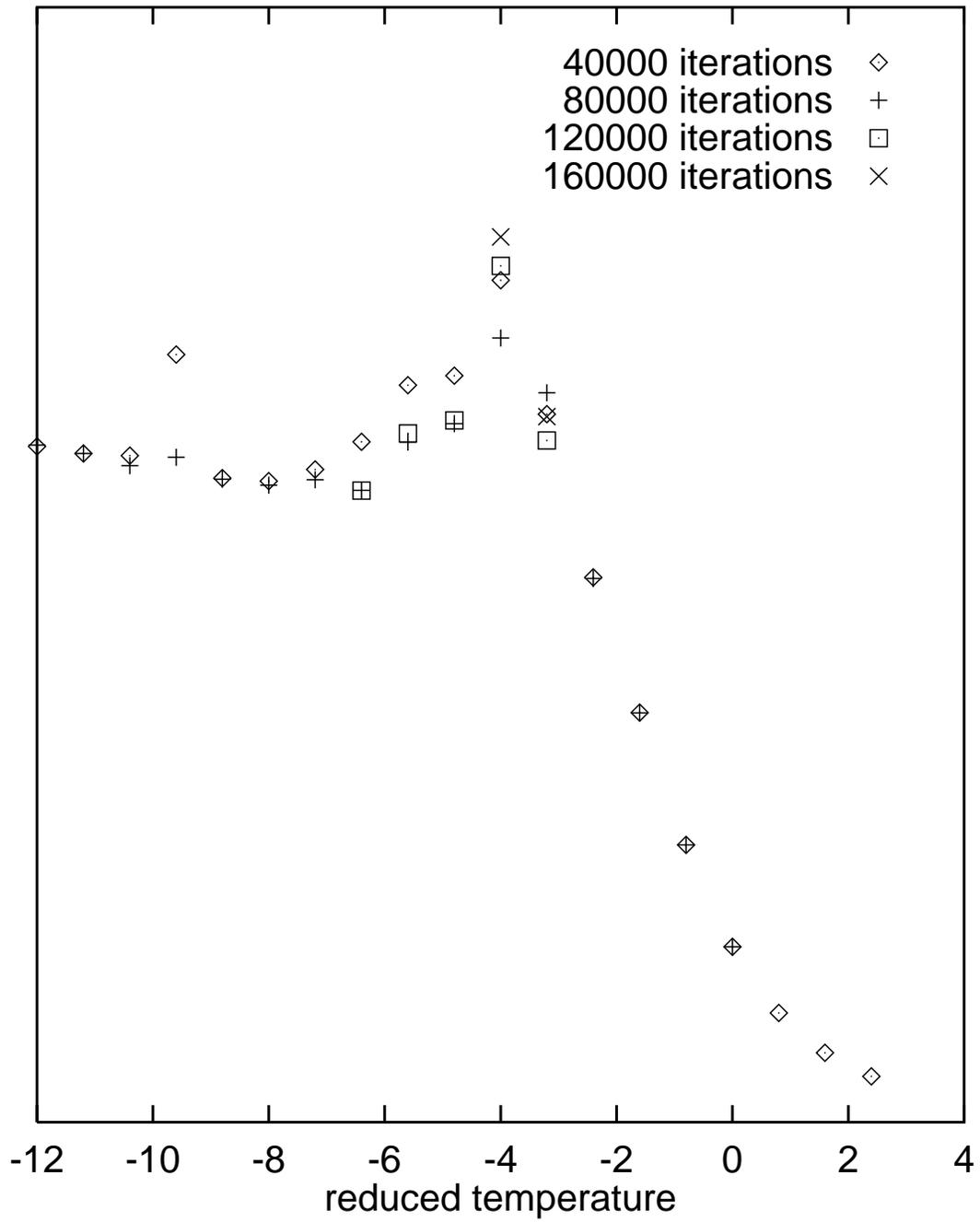
\end{document}